\documentclass[preprint,12pt]{elsarticle}
\usepackage{graphicx} 
\usepackage{caption}
\usepackage{float}
\usepackage{subcaption}
\usepackage{siunitx}
\usepackage{comment}
\usepackage[numbers]{natbib}
\usepackage[section]{placeins}
\usepackage{amssymb}
\usepackage{amsmath}

\makeatletter
\def\ps@pprintTitle{%
  \let\@oddhead\@empty
  \let\@evenhead\@empty
  \let\@oddfoot\@empty
  \let\@evenfoot\@empty}
\makeatother

\begin{document}

\begin{frontmatter}

\title{Advancing Threshold-Inception Modeling for Predictive Simulation of Ionic Wind Fan Performance}

\author[1,3]{Siim Heering}
\author[1]{Juri Volodin}
\author[1]{Vootele Mets}
\author[2]{Rasmus Talviste}
\author[2]{Jüri Raud}
\author[3]{Karl-Eerik Unt}
\author[2]{Indrek Jõgi}
\author[1]{Veronika Zadin}

\affiliation[1]{organization={Institute of Technology, University of Tartu},
            addressline={Nooruse 1},
            city={Tartu},
            postcode={50411},
            country={Estonia}}

\affiliation[2]{organization={Institute of Physics, University of Tartu},
            addressline={W. Ostwaldi 1},
            city={Tartu},
            postcode={50411},
            country={Estonia}}
            
\affiliation[3]{organization={Department of Aeronautical Engineering, Estonian Aviation Academy},
            addressline={Lennu 40},
            city={Reola},
            postcode={61707},
            state={Kambja parish},
            country={Estonia}}
            
\begin{abstract}
    This study investigates the predictive capability of a threshold inception–based multiphysics modeling approach for ionic wind fans by direct comparison with experimental measurements. A wire-to-cylinder electroaerodynamic (EAD) fan with variable electrode spacing is used as a reference system to assess the model’s ability to reproduce airflow characteristics, discharge current, and performance trends under atmospheric conditions.

    Numerical simulations show good qualitative agreement with experimental results across all tested configurations; however, systematic deviations emerge at higher voltages and larger electrode gaps. Analysis of these discrepancies indicates that the commonly adopted assumption of perfectly smooth emitter surfaces can limit model accuracy. Experimental characterization of the emitter wire reveals micro-scale surface protrusions, which locally enhance the electric field and alter corona inception behavior. Incorporating representative surface roughness into the numerical model improves quantitative agreement with measured airflow velocities.

    The results demonstrate that while the threshold inception model provides a robust foundation for EAD fan simulations, electrode surface morphology is a critical factor for reliable prediction. This work advances the validation and refinement of ionic wind fan modeling methodologies and identifies key considerations for the development of more accurate engineering-oriented simulation tools.
\end{abstract}

\begin{keyword}
Electroaerodynamics \sep Electrohydrodynamics \sep Ionic wind \sep Multiphysics simulation \sep Corona discharge
\sep Threshold inception \sep Surface roughness \sep Drift-diffusion
\end{keyword}

\end{frontmatter}

\section{Introduction}

For decades, the generation and control of airflow have relied predominantly on mechanically driven devices such as propellers, fans, and rotors. While mature and effective, these systems inherently involve moving parts, leading to acoustic noise, mechanical wear, and increased system complexity. In recent years, growing interest in quieter, more robust, and energy-efficient technologies has motivated renewed attention toward alternative flow-generation mechanisms. One such approach is based on electroaerodynamic (EAD), or equivalently electrohydrodynamic (EHD), momentum transfer, which enables airflow generation without moving mechanical components \cite{Park2018TheForce}.

The EAD effect arises when a sufficiently strong electric field ionizes the surrounding gas, typically air, and accelerates the resulting charged particles toward an oppositely biased electrode. Through frequent collisions, these ions transfer momentum to neutral molecules, producing a bulk airflow commonly referred to as ionic wind. Depending on the configuration and operating regime, this mechanism has found application across a wide range of technologies, including electrostatic precipitators for air purification, EHD pumps, active cooling of electronic components, and flow manipulation for aerodynamic control \cite{Jaworek2025RecentPaper}, \cite{Kriegseis2016TowardsControl}, \cite{Wang2017ExperimentalWind}, \cite{Qu2017RecentApplications}, \cite{Fu2007AirflowActuators}. Furthermore, EAD has been explored as a propulsion mechanism \cite{Leng2024OnAnalysis}, \cite{Cheng2024EHDChannel}, \cite{Iranshahi2024ElectrohydrodynamicsPerspectives}, \cite{Xu2018FlightPropulsion}, as well as a means of influencing aerodynamic boundary layers to reduce skin-friction drag\cite{Leger2002EffectPlate}, \cite{Moreau2006Effect25m/s}, \cite{El-Khabiry1997DragFlow}. In such aerodynamic applications, EAD-induced momentum addition interacts with an external freestream, enabling local flow acceleration and boundary-layer modification rather than bulk air transport. It must be noted, that airflow generation and aerodynamic flow control should not be regarded as competing functions, but rather as complementary manifestations of the same underlying EAD momentum transfer process.

The  benefits of EAD-based flow control are closely linked to the broader demand for improved energy efficiency across multiple sectors. For example, even modest reductions in profile drag can lead to significant cumulative fuel savings in commercial aviation while improved aerodynamic efficiency is also a key objective in wind energy applications, where it enables increased power extraction and a broader operational envelope \cite{NabhaniLarge-ScaleJets}. Previous studies have shown that EAD conversion efficiency increases with air velocity, reaching higher values at elevated freestream speeds \cite{Xu2018FlightPropulsion}, \cite{Pogorelov1986EffectDevice}, further highlighting the synergy between ionic wind generation and externally driven flows. The need for improving the aforementioned efficiency brings out the necessity for a robust and reliable engineering tool that would provide reliable and fast results.

Regardless of the fact the EAD has vast application potential and amount of research being done so far, the current technological readiness level is still in early stages due to several reasons. Most noteworthy are low conversion efficiency (electrical to mechanical energy), longevity issues caused by the high electric field and its interaction with the electrode, need for compact and safe high voltage power supplies \cite{Iranshahi2024ElectrohydrodynamicsPerspectives}, \cite{CogollodeCadiz2021MaterialsDischarge}.

Despite decades of experimental and numerical research, the reliable prediction of ionic wind performance remains challenging. This difficulty stems from the inherently multiphysics nature of EAD systems, which involve strong coupling between electrostatics, charge transport, and fluid dynamics \cite{Narvaez-Munoz2024ComputationalApplications}, \cite{Yan2022CouplingActuator}, \cite{Gao2025ReviewApplications}. In particular, corona discharge phenomena introduce nonlinear feedback mechanisms: the buildup of space charge alters the local electric field, which in turn affects further ionization and momentum transfer. Accurately capturing these interactions is essential for predictive modeling, especially when numerical tools are intended for engineering-oriented design and optimization rather than qualitative trend analysis.

There are numerous numerical models describing EAD and one of them is a drift-diffusion model, which provides balance between fidelity and robustness. However, it has its limitations regarding ionization and taking into account effects like local field enhancements. This is where threshold-inception model comes in that focuses exactly on the aforementioned shortcomings of the drift-diffusion model. Threshold-inception model offers a computationally efficient framework for simulating corona discharges by defining ionization onset based on local electric-field criteria under given atmospheric conditions \cite{Mikropoulos2015ThresholdConditions}. This model has demonstrated promising results for simplified electrode arrangements, but their quantitative predictive capability for practical EAD fan configurations has not yet been comprehensively validated against experimental performance data.

A representative example is the comprehensive numerical study by Bedolla et al. \cite{Bedolla2017AltitudeLifter}, who investigated the altitude dependence of electrohydrodynamic thrust. Their model accounts for variations in air density, temperature, pressure, and humidity, and successfully reproduces experimentally observed trends in thrust reduction with altitude. Corona inception in that work is enforced through an empirical surface-field criterion, with the surface charge density adjusted to satisfy a steady-state condition consistent with Kaptsov-type assumptions. While this approach is well suited for studying environmental dependencies and steady EAD flow behavior, it does not explicitly resolve the transient buildup and saturation of space charge in the ionization region, nor does it dynamically couple ionization onset to evolving electrostatic shielding. This is where threshold-inception-based approach promises to help, in which ionization is introduced conditionally once the local electric field exceeds a prescribed inception threshold. In such formulations, the development of space charge and the resulting electrostatic shielding emerge naturally as part of the solution, rather than being imposed a priori. This framework offers a computationally efficient means of representing corona discharge physics while retaining a physically motivated description of ionization onset under varying atmospheric conditions. Threshold-inception models have shown promising results in simplified geometries, but their quantitative predictive capability for practical EAD devices has not yet been comprehensively validated against experimental performance data.

In addition to ionization modeling, geometric realism plays a crucial role in predictive accuracy. Many numerical studies assume perfectly smooth electrode surfaces, an idealization that simplifies field calculations but does not reflect manufactured components. In practice, metallic emitter electrodes exhibit micro-scale surface roughness and defects arising from fabrication processes or material degradation \cite{CogollodeCadiz2021MaterialsDischarge}. Such features can locally enhance the electric field, modify effective corona inception conditions, and alter the spatial distribution of space charge. Neglecting these effects may therefore lead to systematic discrepancies between simulated and measured performance, particularly at higher voltages or larger electrode gaps.

This brings us to the matter of physical EAD configurations. As the technology is still maturing, then different electrode arrangements are being investigated to better understand the characteristics based on the function of the system. Given some examples of electrode arrangements like wire-to-cylinder \cite{Moreau2013ElectrohydrodynamicPressure}, wire-to-plate \cite{Leger2002EffectPlate}, needle-to-cylinder \cite{Lee2015ParallelRate}, needle-to-mesh \cite{Moreau2008EnhancingDischarges} and wire-to-airfoil \cite{Grosse2024ElectroaerodynamicDischarge} shows the mechanical simplicity of the system and point out the different characteristics. Furthermore, significant research effort has been devoted to integrating electrodes directly into surfaces using printed circuit board technologies or specialized laminates, particularly in the context of dielectric-barrier-discharge (DBD) actuators \cite{Leger2002EffectPlate}, \cite{Lee2015ParallelRate}, \cite{Sato2019SuccessivelyOperation}, \cite{Sato2021DevelopmentInk}. Among these configurations, the wire-to-cylinder arrangement remains one of the most extensively studied versions and provides a well-defined geometry for combined experimental and numerical investigations. 

The present study addresses the limitations of current drift-diffusion model by developing and validating a fully coupled electroaerodynamic (EAD) modeling framework that combines drift–diffusion charge transport with a threshold-inception-based description of corona discharge. In this approach, ionization is introduced dynamically once the local electric field exceeds the inception threshold, allowing the space-charge development and electrostatic shielding effects to emerge naturally from the solution.

The model is applied to a wire-to-cylinder ionic wind fan configuration, selected as a well-defined and scalable reference system suitable for both experimental measurement and numerical analysis. Direct comparison with experimental data is used to assess the model’s ability to reproduce airflow characteristics, discharge current, and performance trends across a range of operating conditions.

As a second objective, the influence of representative electrode surface roughness is examined by incorporating micro-scale protrusions into the emitter geometry. This analysis aims to quantify the impact of realistic surface morphology on corona inception behavior and overall predictive accuracy. Rather than proposing a universal correction, the study seeks to identify and demonstrate the significance of electrode surface effects as a previously under-represented factor in engineering-oriented EAD simulations.

\section{Methods}

\subsection{Governing equations}

Several governing equations describe the ionic wind phenomenon. For describing the flow of air Navier-Stokes's equations are being used, which define the conservation of mass (Eq 1), momentum (Eq 2), viscous stress tensor (Eq 3) and strain rate tensor (Eq 4), while considering air to be a Newtonian fluid
\begin{equation} \label{eq1}
\nabla\cdot v = 0
\end{equation}

\begin{equation} \label{eq2}
\rho\frac{\partial v}{\partial t} + \rho(v \cdot \nabla)v=\nabla \cdot [-pI+K]+ F
\end{equation}

\begin{equation} \label{eq3}
K = 2\mu S-\frac{2}{3}\mu (\nabla \cdot v)I
\end{equation}

\begin{equation} \label{eq4}
S = \frac{1}{2}(\nabla v + (\nabla v)^ T)
\end{equation}
where 
\(v\) is considered here as a velocity, \(\rho\) being density of the fluid, \(p\) being pressure, \(I\) being an identity matrix, \(K\) being viscous stress tensor, \(F\) being volume force vector, \(\mu\) being dynamic viscosity of the medium and \(S\) being the strain rate tensor.

In the case of ionic wind Eq 5 can be written based on Eq 2 as
\begin{equation} \label{eq5}
\rho_{air}\frac{\partial v}{\partial t} + \rho_{air}( v\cdot\nabla) v = -\nabla p + \mu\nabla^2 v + \rho_{sc} E,
\end{equation}
where \(\rho_{air}\) is density of air, \(v\) velocity vector, \(p\) is pressure, \(\rho_{sc}\) is space charge density and \(E\) is electric field strength. First term on the right side of equation 5 represents pressure gradient forcing the air to move in the direction of most significant change of pressure. Second term represents the diffusion term that is determined by the dynamic viscosity factor. Third term (\(\rho_{sc} E\)) is considered to be the external force affected by space charge density and electric field strength.

From the electrostatics perspective relevant equations are (Eq 6,7,8)
\begin{equation} \label{eq6}
E=-\nabla V
\end{equation}
\begin{equation} \label{eq7}
\nabla \cdot E=\frac{\rho_{sc}}{\varepsilon}
\end{equation}
\begin{equation} \label{eq8}
\nabla^2 V=-\frac{\rho_{sc}}{\varepsilon_0\varepsilon_r},
\end{equation}
where \(\varepsilon_0\) is a dielectric permittivity of free space and \(\varepsilon_r\) is relative permittivity of air. 

Electric current flow has three contributors – conduction, convection and diffusion. Conduction as motion of ions relative to the airflow, convection as migration of charged particles within the airflow and diffusion as a process caused by the difference of volume charge densities. Combining these three factors gives a total current flux J (Eq 9)
\begin{equation} \label{eq9}
J = \mu_E E \rho_{sc} + \rho_{sc} v - D\nabla \rho_{sc}.
\end{equation}
The first term on the right side of the equation is the conduction term, where \(\mu_E\) is the mobility of charged particles in air. The second term in the equation is the convection term and it is a product of space charge density and velocity vector. Final term is the diffusion term where D is the diffusion coefficient of ions. 

Assumption can be made that no ions are being generated in the drift zone, so a continuity equation can be formed (Eq 10)
\begin{equation} \label{eq10}
\rho_{sc}\frac{\partial v}{\partial t} +\nabla\cdot J = 0.
\end{equation}

In the computational model the electric current flux formula is being represented by the stabilized convection equation (Eq 11)
\begin{equation} \label{eq11}
d_a \frac{\partial \rho_{sc}}{\partial t} + \nabla \cdot( - D\nabla \rho_{sc} + \mu_E E \rho_{sc} + \rho_{sc} v)= f.
\end{equation}

The model is based on \cite{Bedolla2017AltitudeLifter} where the effect of air pressure, humidity and temperature were considered. However, Bedolla's model doesn't dynamically take into account the ionization inception, so the threshold-inception model, proposed by \cite{Mikropoulos2015ThresholdConditions} (Eq 12) determines the electric field strength on the surface of the emitter wire.

\begin{equation} \label{eq12}
E_s = E_0[1+\frac{0.568}{(\delta r_0)^{0.34}}]\cdot(K_H)^{-0.2},
\end{equation}

where \(E_0\) is the breakdown electric field in air, \(\delta\) is relative air density (compared with standard air density) and \(r_0\) is the radius of emitter wire in cm. \(K_H\) is humidity correction factor in terms of absolute humidity \(H\) (\(\frac{g}{m^3}\)) and is expressed as

\begin{equation} \label{eq13}
K_H = 1 + 0.012H^{0.624\delta^{0.00624}}.
\end{equation}

The coupling between the electric field and the flow field is introduced through Gauss’s law (Eq. 7). The volume force acting on the fluid is then obtained by combining the space charge density with the electric field (Eq 14)

\begin{equation} \label{eq14}
F = \rho_{sc} E.
\end{equation}

\subsection{Experimental setup}

The schematic of the experimental setup is shown in Figure 1 and image of it in Figure 2. Rectangular cross-sectional geometry was studied in the experiments. The emitter electrode consisted of 7 thin wires with diameter of 0.06 mm and 5.5 mm distance between the wires. 
The collector electrode consisted of 7 tubes with diameter of 2 mm and placed in alignment with the wires.
The vertical distance d between the collector and emitter was varied between 5, 10 and 15 mm. The positive corona discharge was powered by a DC power supply by applying a positive voltage to thin wires. The applied voltage was varied between 4 and 14 kV and measured with a voltmeter. The flow speed of air from the corona discharge was measured with a Testo 425 flowmeter placed 150 mm downstream from the grounded collector electrode. Discharge current was measured with a microammeter. Experiments were carried out in laboratory air at atmospheric pressure at 23°C and 46\% relative humidity. The day-to-day variations were +/- 1°C and +/- 3\% within the experimental period.
Averaged experiment results can be found in Table 1. The variations in temperature and humidity were numerically verified to have insignificant effect to the output speed.

\begin{figure}[H]
    \centering
    \includegraphics[width=\textwidth]{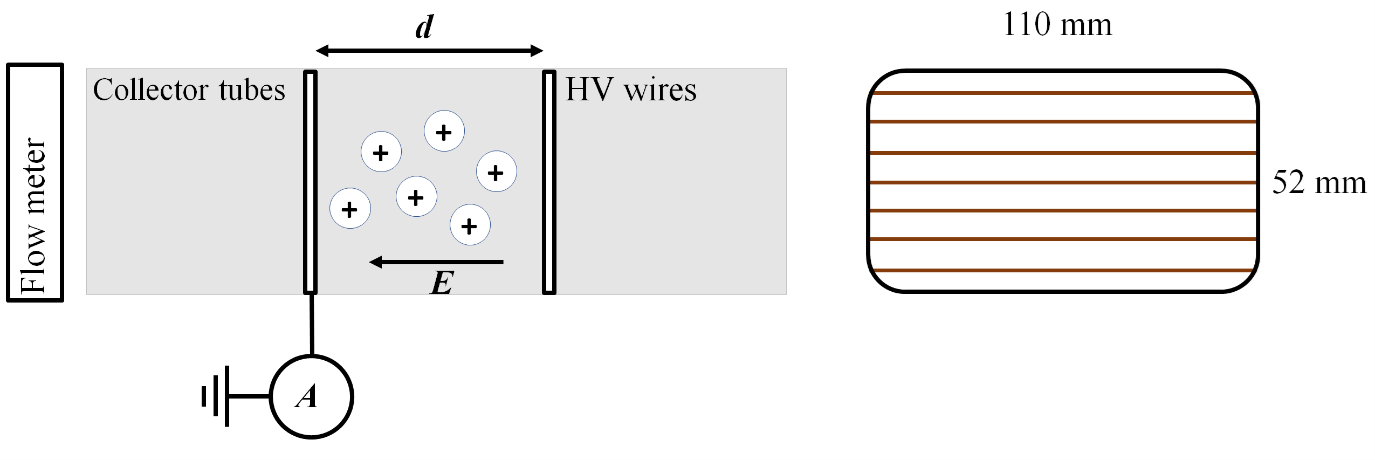}
    \label{fig:f1}
    \caption{Experimental setup and cross-section geometries used}
\end{figure}

As can be seen from Figure 2, the fan has been placed inside a plastic duct and 3D-printed structure is supporting the collector and emitter wires. Distance between the wires can be altered by changing the distance bushings between the two frames. 

\begin{figure}[H]
    \centering
    \includegraphics[width=\textwidth]{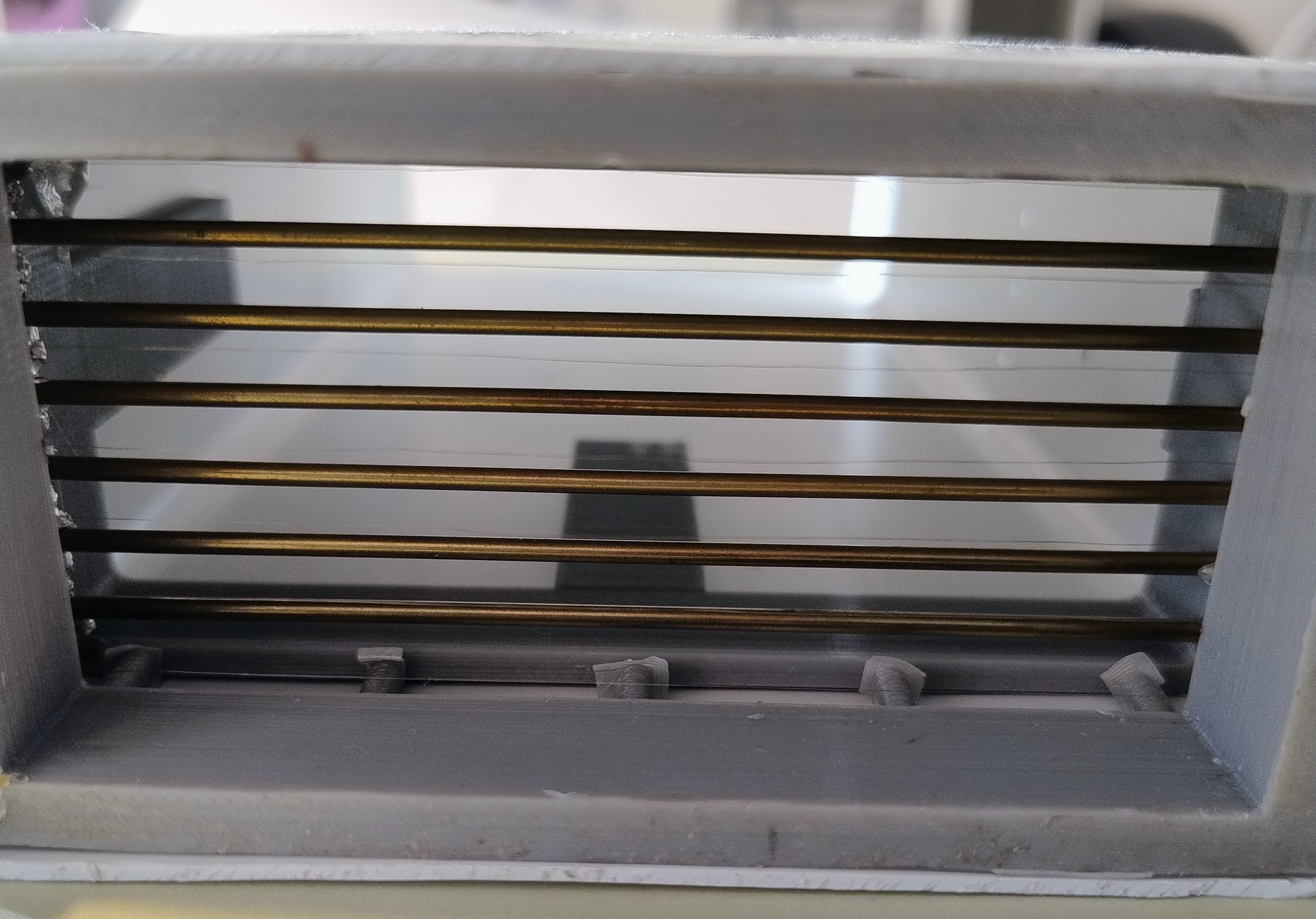}
    \label{fig:f1}
    \caption{Image of the experimental setup}
\end{figure}

\subsection{Computational model setup}

The computational model was implemented in COMSOL Multiphysics 6.2 using a fully coupled multiphysics formulation. The governing equations were realized through the Laminar Flow, Electrostatics, Stabilized Convection–Diffusion Equation, and Distributed ODE interfaces. Owing to the relatively low induced air velocities, the flow was assumed to be incompressible and laminar.

Electroaerodynamic coupling was achieved by introducing a volumetric body force in the laminar flow module, representing momentum transfer from charged particles to the neutral air. The magnitude of this force depends on the local space charge density and electric field strength. Charge transport in the drift region was modeled using a stabilized convection–diffusion formulation that accounts for ionic conduction, convection by the airflow, and diffusion, consistent with the governing equations presented in Section 3.1.

The computational domain reproduces the experimental duct geometry. No-slip boundary conditions were applied at all solid walls. To account for the inertia of the surrounding air mass and to enable physically meaningful transient development of the induced flow, an auxiliary airbox domain was introduced downstream of the duct. The outlet velocity profile of the duct was coupled to the airbox using a linear extrusion approach, allowing the induced flow to develop against a finite air volume rather than an artificial pressure boundary. Total pressure boundary conditions were applied at the inlet and outlet of the airbox, ensuring that both static and dynamic pressure components are consistently accounted for.

The numerical mesh consisted of unstructured triangular elements with boundary-layer refinement near solid surfaces and electrodes. Depending on the electrode spacing, the total number of degrees of freedom was approximately one million. The nonlinear system was solved using a fully coupled constant-Newton approach, while the linearized systems were handled using a direct solver (MUMPS). This configuration ensured numerical robustness and convergence across the full range of simulated operating conditions.

\begin{figure}[H]
    \centering
    \includegraphics[width=1\linewidth]{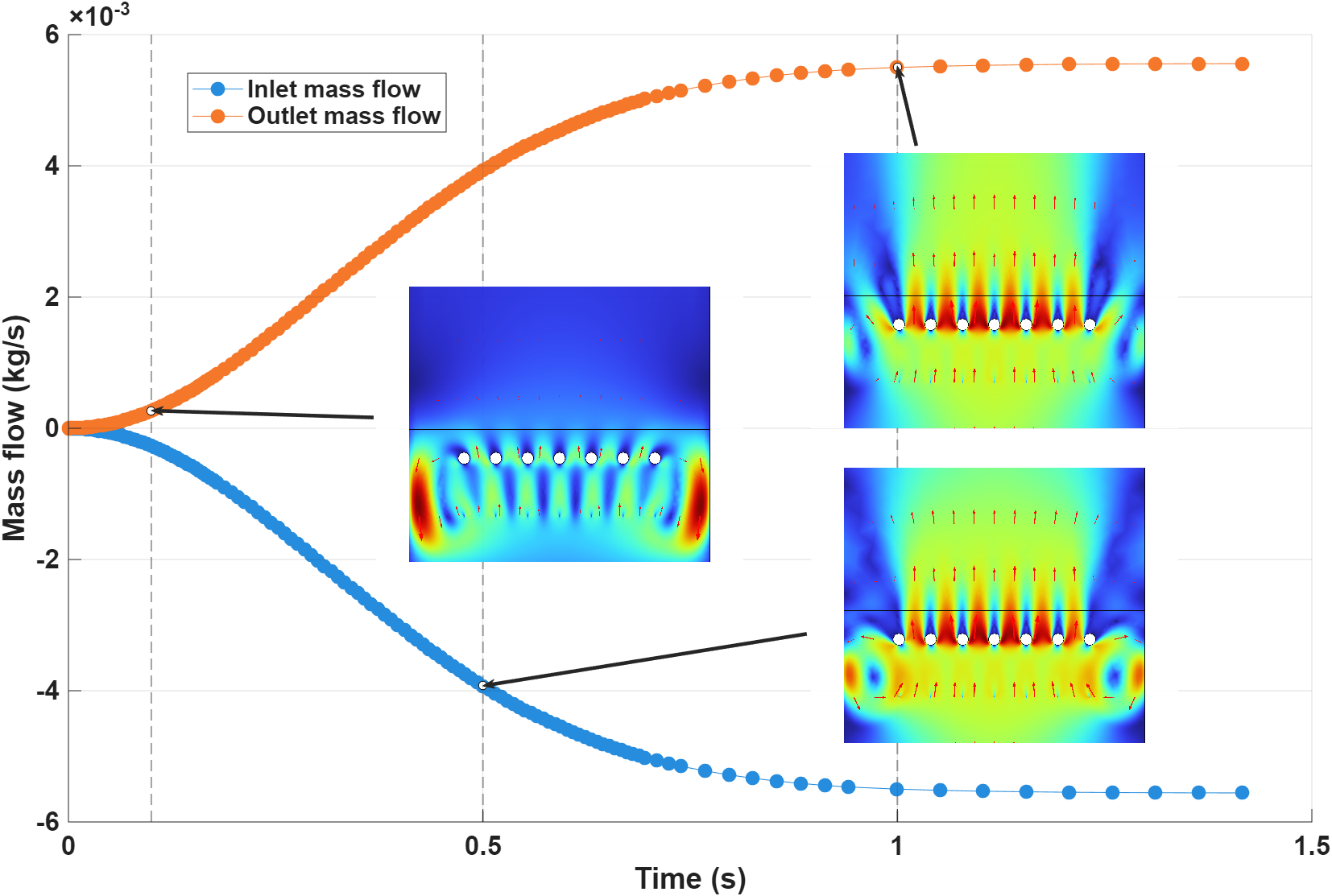}
    \caption{Mass flow graph with visualized air flow distribution}
    \label{fig:f5}
\end{figure}

Figure 3 shows the mass flow in and out of the duct and airbox respectfully. At used conditions the mass flows stabilize in approximately 1 s. As can be seen then flows match between duct and airbox indicating that duct and airbox boundaries are linked together successfully. Figure 4 visualizes the airflow velocity distribution during the stabilized mass flow period.

\begin{figure}[H]
    \centering
    \includegraphics[width=1\linewidth]{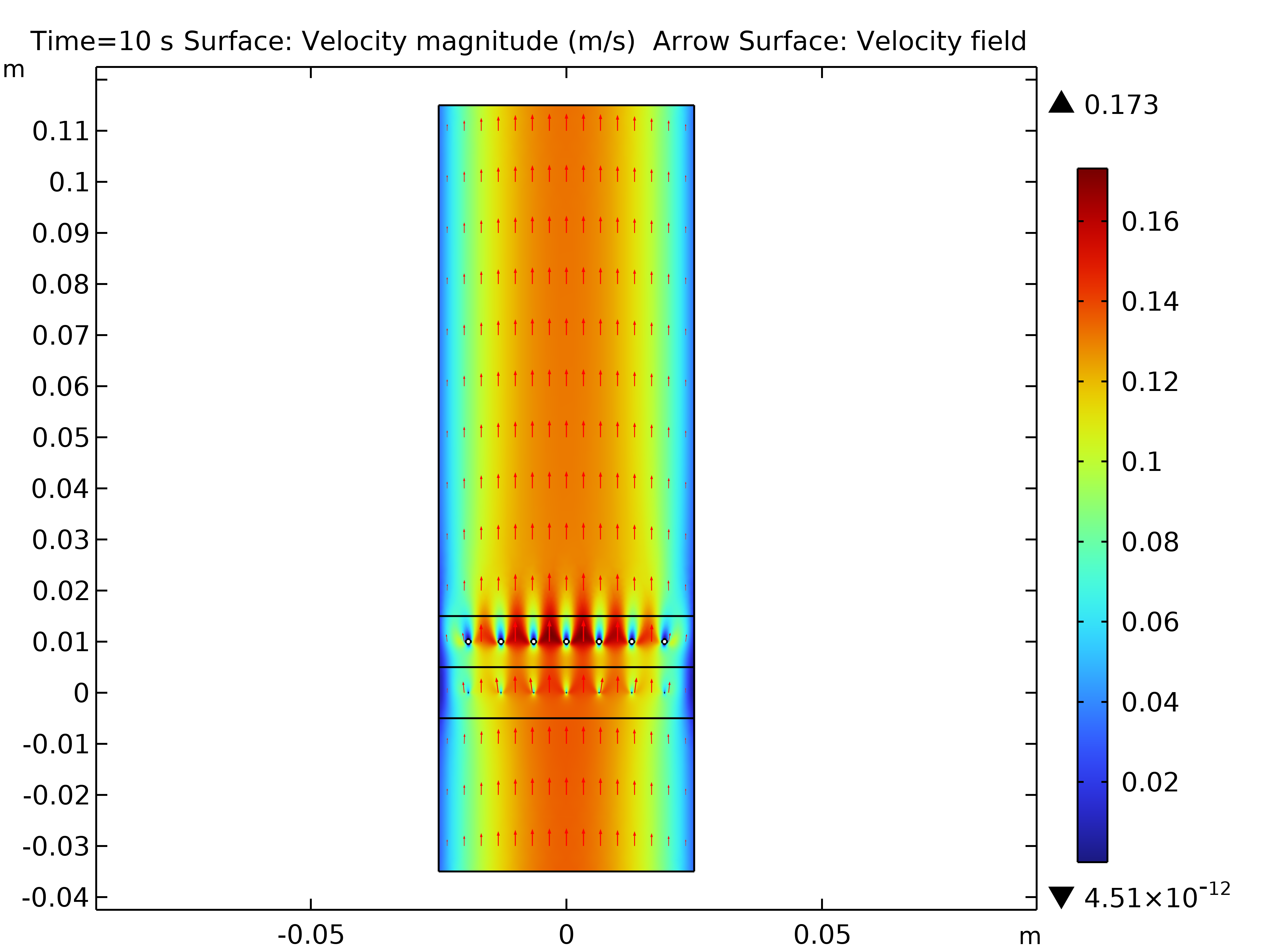}
    \caption{Airflow velocity in duct}
    \label{fig:f6}
\end{figure}

The main geometric, physical, and electrical parameters used in the simulations are summarized in Table 1. The boundary conditions applied to the electrostatic (Eq. 6–8), charge transport (Eq. 11), and fluid-flow equations (Eq. 2) are summarized in Table 2. These conditions are consistent with the governing equations presented in Section 3.1 and represent the physical operation of the experimental system, where the emitter wires are maintained at the applied high voltage, the collector electrodes are grounded, and no-slip conditions are imposed at all solid boundaries.

\begin{table}[H]
    \centering
    \begin{tabular}{c|c}
        \textbf{Parameter} & \textbf{Value} \\
        \hline
         Emitter to collector distance, mm & 5, 10, 15 \\
         Applied voltage, kV & 6 - 15 \\
         Breakdown electric strength in air, V/m & $3.31 \times 10^6$ \\
         Ion mobility coefficient, m$^2$/(V $\cdot$ s) & $2 \times 10^{-4}$ \\
         Electric field strength at the surface of the emitter, V/m & $3.66 \times 10^6$ \\
         Duct width, mm & 50 \\
         Duct height, mm & 100 \\
         Emitter wire radius, mm & 0.03 \\
         Collector rod radius, mm & 0.5 \\
         Distance between electrode pairs, mm & 5.5 \\
         Number of electrode pairs & 7 \\
    \end{tabular}
    \caption{Main parameters used in the simulation}
    \label{tab:placeholder}
\end{table}

\begin{table}[H]
    \resizebox{\linewidth}{!}{
    \centering
    \begin{tabular}{c|c|c|c}
        Boundary & Electrostatics & Convection-Diffusion & Fluid dynamics \\
        \hline
         Emitter wires & \(V = V_e\) & \(\rho_{sc}\) from inception ODE & \(v = 0\) \\
         Collector rods & Grounded, \(V = 0\) & \(\rho_{sc} = 0\) & \(v = 0\) \\
         Channel walls & Zero charge, \(n\cdot D = 0\) & \(-n \cdot (-D\nabla \rho_{sc}) = 0\) & \(v = 0\) \\
         Channel inlet & Zero charge, \(n\cdot D = 0\) & \(\rho_{sc} = 0\) & \(p = p_0 - \frac{1}{2}\rho_{air}|v|^2\) \\
         Channel outlet & Zero charge, \(n\cdot D = 0\) & \(-n \cdot (-D\nabla \rho_{sc}) = 0\) & \(p = p_0 - \frac{1}{2}\rho_{air}|v|^2\) \\
         Air domain in duct & Eq 14 & Eq 11 & Eq 2 with \(F=\rho_{sc}E\) \\
    \end{tabular}
    }
    \caption{Boundary conditions in the simulation setup}
    \label{tab:placeholder}
\end{table}

\subsubsection{Coupling between flow field and charge density}

The threshold-inception formulation is not implemented as a standalone logical condition, but as a dynamically coupled boundary mechanism within the drift–diffusion framework.

A Distributed ODE is solved on the emitter surface to describe the evolution of surface charge density during voltage ramp-up. At each time step, the local electric field strength at the emitter surface \(E_p\) is compared with the inception threshold \(E_s\). When \(E_p < E_s\), no ionization occurs and the surface charge remains constant. Once \(E_p>E_s\), the surface charge density increases proportionally to the excess field strength.

Importantly, the surface charge density determined by the inception ODE serves as a boundary condition for the drift–diffusion equation governing space charge transport in the gas domain (Eq. 11). In this way, the inception model directly regulates the injection of charge into the computational domain.

The resulting space charge density \(\rho_{sc}\) enters Poisson’s equation (Eq. 8), modifying the electric potential and consequently the electric field distribution. The altered electric field then feeds back into the inception condition through the updated surface field \(E_p\). 

This establishes a closed nonlinear feedback loop:

surface field → surface charge growth → space charge transport → electric field redistribution → surface field regulation. This has also been shown schematically in Figure 5.

As space charge accumulates near the emitter, electrostatic shielding reduces the local electric field strength until \(E_p=E_s\), at which point the surface charge density stabilizes and a steady corona regime is reached.

Through this mechanism, the threshold-inception model is fully embedded within the drift–diffusion–Poisson system rather than imposed as an external constraint. The electrohydrodynamic body force (Eq. 14) then couples the charge distribution to the flow field, completing the multiphysics interaction.

\begin{figure}[H]
    \centering
    \includegraphics[width=1\linewidth]{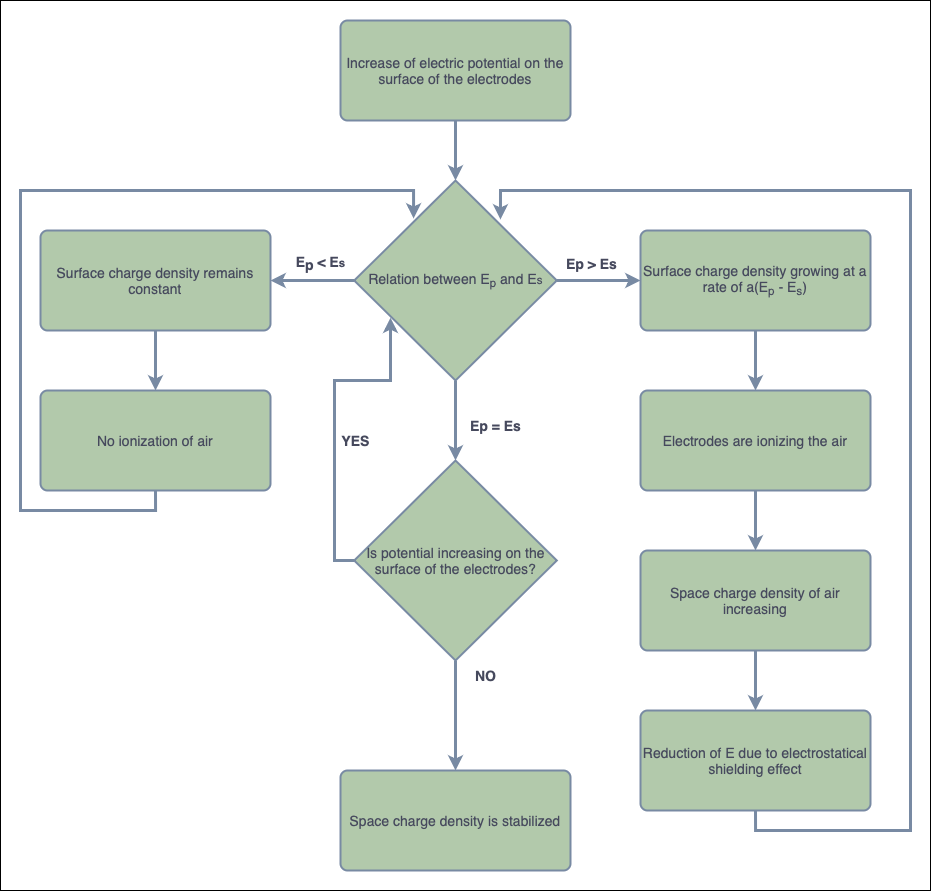}
    \caption{Schematic explaining the coupling between surface charge density, space charge density and electric field used in the simulation}
    \label{fig:f5}
\end{figure}

\section{Results}

\subsection{Experiment results}

The experimental setup performance was evaluated based on measured parameters (Fig 6). 

\begin{figure}[H]
    \begin{subfigure}[b]{0.32\textwidth}
        \includegraphics[width=\textwidth]{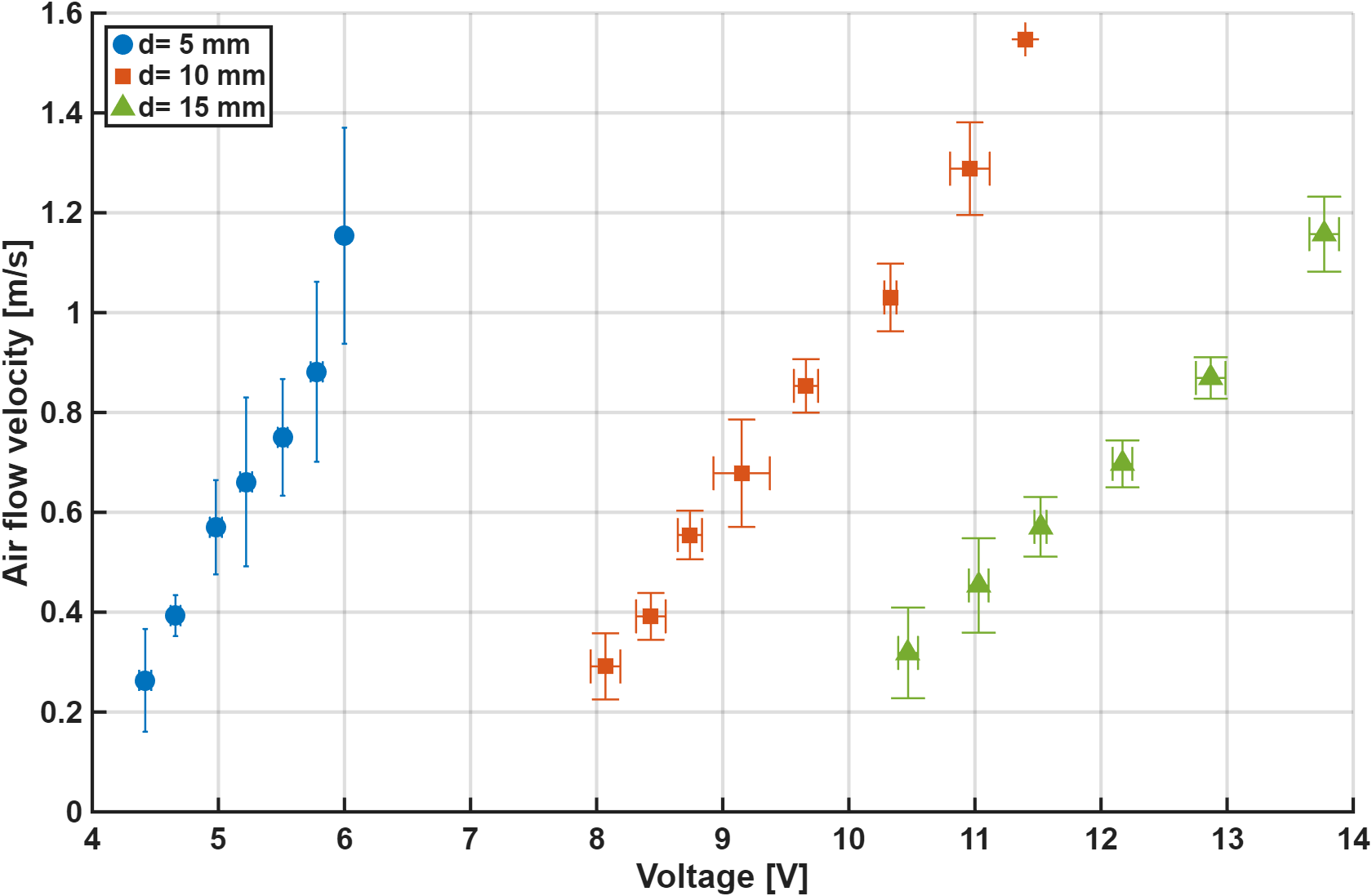}
        \caption{}
    \end{subfigure}
    \hfill
    \begin{subfigure}[b]{0.32\textwidth}
        \includegraphics[width=\textwidth]{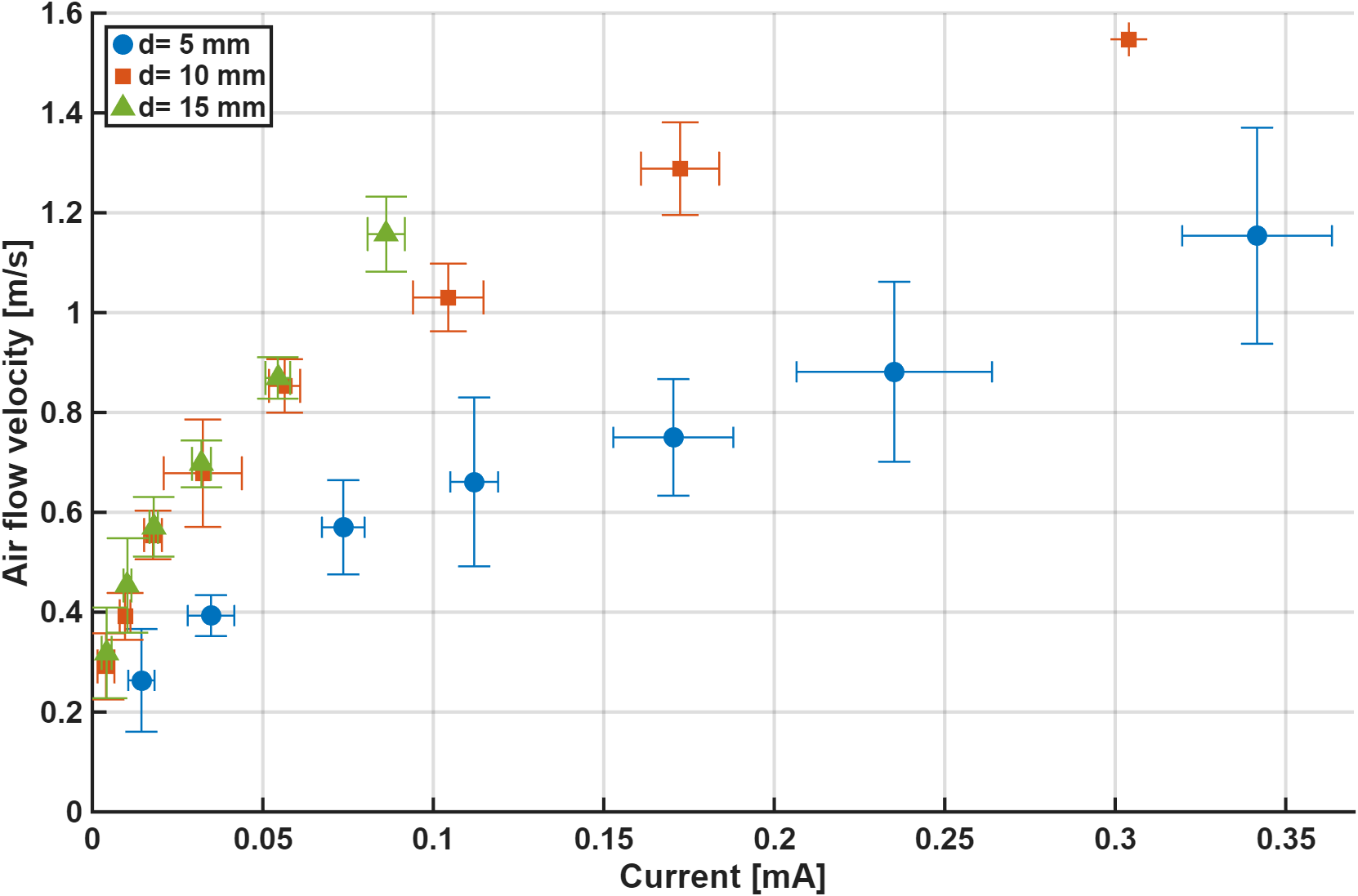}
        \caption{}
    \end{subfigure}
    \hfill
    \begin{subfigure}[b]{0.32\textwidth}
        \includegraphics[width=\textwidth]{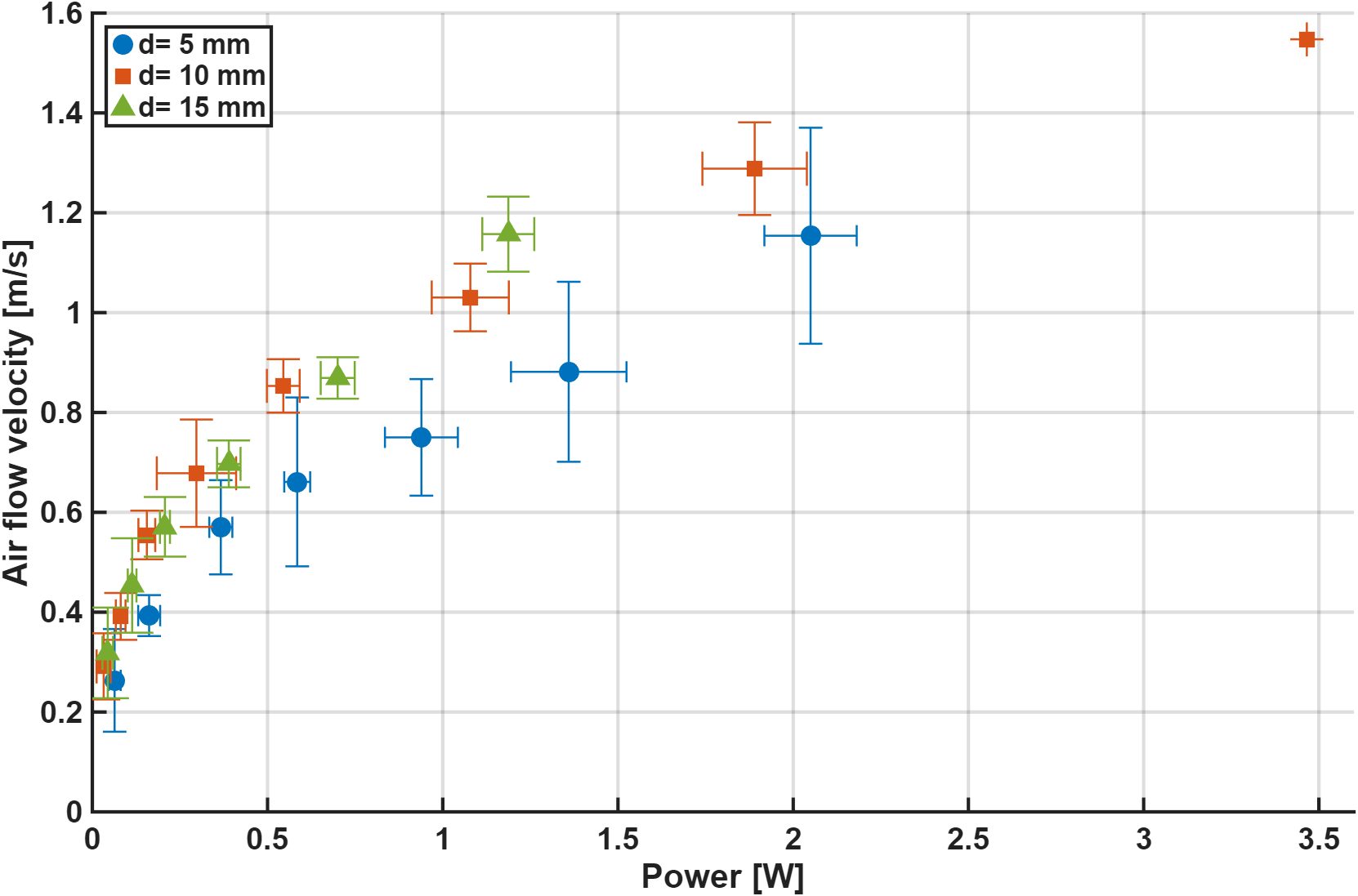}
        \caption{}
    \end{subfigure}
    \caption{Experimental data and comparison of airflow, current and power with respect to electrode gap}
    \label{fig:f16}
\end{figure}

Experiment's highest output velocity of 1.55 m/s occurred at a 10 mm electrode gap, 11.4 kV emitter voltage, and 0.3 mA discharge current. With given atmospheric conditions the calculated air density is \(\rho = 1.186 kg/m^3\) and the mass flow \(\dot{m}\) can be found using Eq 15

\begin{equation}
    \dot{m} = \rho\, A_{\mathrm{exit}}\, v = 10.5 g/s,
\end{equation}

where \(A_{\mathrm{exit}}=57.2 cm^2\) being the cross-section of the duct. Approximate thrust output can be found using Eq 16

\begin{equation}
    T \approx \dot{m}\Delta v \approx 16.3 mN
\end{equation}

For finding the efficiency of the setup the aerodynamical flow power needs to be found. This can be found from Eq 17.

\begin{equation}
    P_{mech} = \frac{1}{2}\dot{m}v^2 = 12.6 mW
\end{equation}

With Eq 18 the electrical power can be found

\begin{equation}
    P_{elec} = V\cdot I = 3.47W
\end{equation}

By comparing the mechanical and electrical power the efficiency of the system can be found using Eq 19.

\begin{equation}
    \eta = \frac{P_{mech}}{P_{elec}} = 0.36\%
\end{equation}

As in similar studies,  such as \cite{Ramadhan2017NumericalChannel} and \cite{Moreau2008EnhancingDischarges} ,the calculated efficiency is low, but within reasonable range (between 0.2 and 2 \%). One of the main contributor for low efficiency value is most probably the larger diameter of emitter wires that were used in the experiment. 

Thrust-per-power was found to be \(\frac{T}{P_{elec}}=4.7 mN/W\) and thrust-per-current \(\frac{T}{I}=54.3N/A\). These values show that the EAD fan used in the experiment as the reference system could improve in terms of efficiency and output power, however it still shows that the experimental setup is suitable to be used as reference system to validate the numerical method.

The emitter wire, made of Nichrome, used in the experiment was scanned under SEM (Fig. 7) and the results showed relatively good surface roughness with some protrusions running along the wire. The measured height of the protrusions was around 2 \(\mu m\) and this value was taken as the height of the protrusion in the simulation. It is assumed that the defects running along the length of the wire surface are caused by the manufacturing process in which the wire is being drawn through different sizes of dyes to reach the required dimension. 

\begin{figure}[H]
    \begin{subfigure}[b]{0.56\textwidth}
        \includegraphics[width=\textwidth]{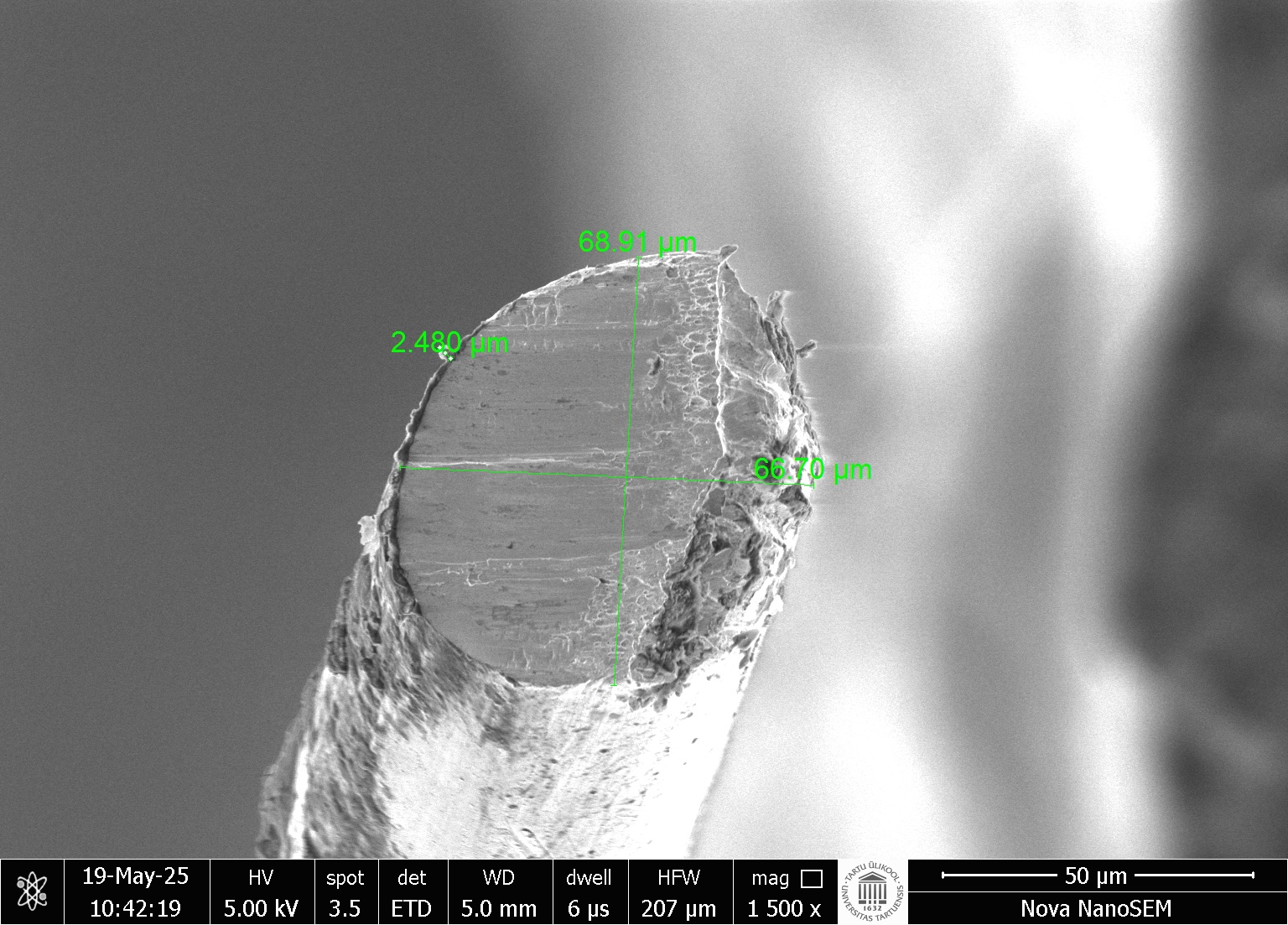}
        \caption{}
        \label{fig:f3}
    \end{subfigure}
    \hfill
    \begin{subfigure}[b]{0.44\textwidth}
        \includegraphics[width=\textwidth]{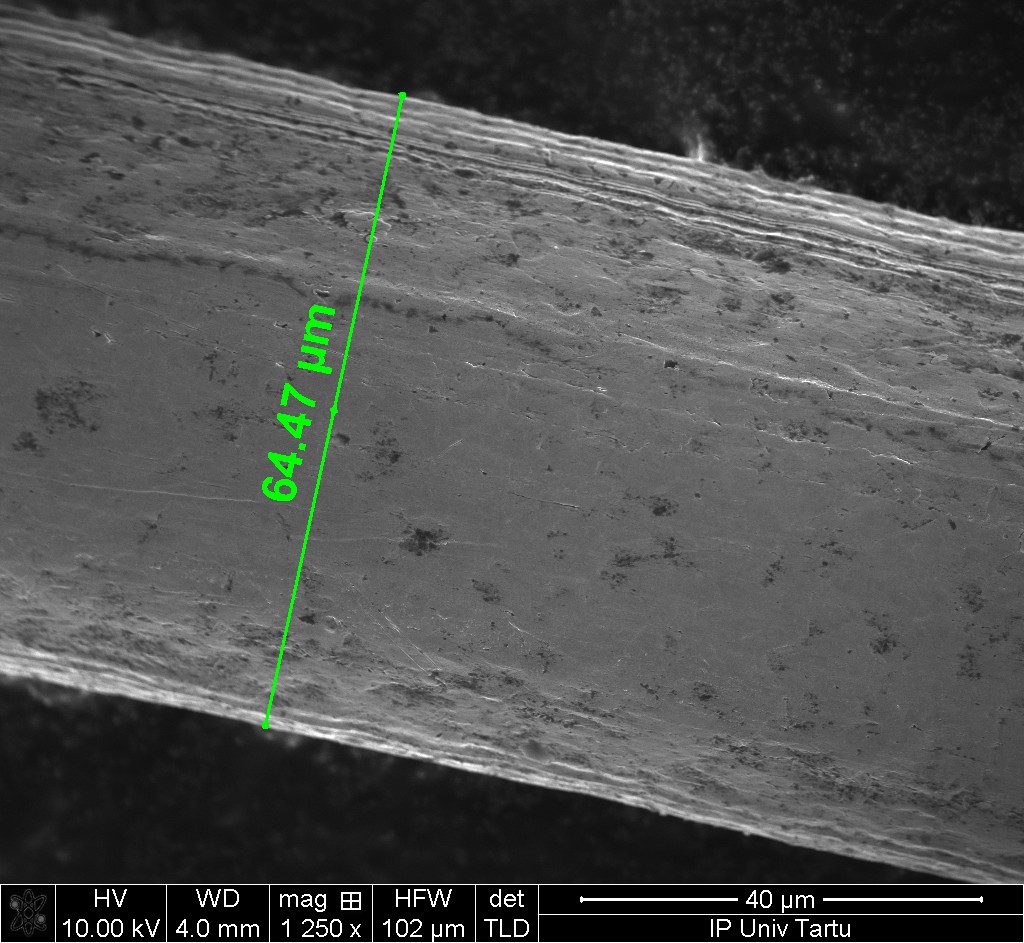}
        \caption{}
        \label{fig:f4}
    \end{subfigure}
    \caption{SEM images of emitter wire's cross-section and side}
\end{figure}

\subsection{Simulation results}

Simulations were conducted for electrode spacings of 5, 10, and 15 mm using the same applied voltage range as in the experiments. Overall, the numerical model reproduces the experimentally observed trends in airflow generation with good qualitative agreement and reasonable quantitative accuracy (Fig. 8a). In particular, the model correctly captures the increase in induced airspeed with applied voltage and the dependence of performance on electrode spacing.

However, for larger electrode gaps the predicted voltage-velocity slopes diverge from the experiment data. This behavior suggests that certain physical effects influencing ionization and momentum transfer are not fully represented in the baseline model configuration.

\begin{figure}[H]
    \begin{subfigure}[b]{0.48\textwidth}
        \includegraphics[width=\textwidth]{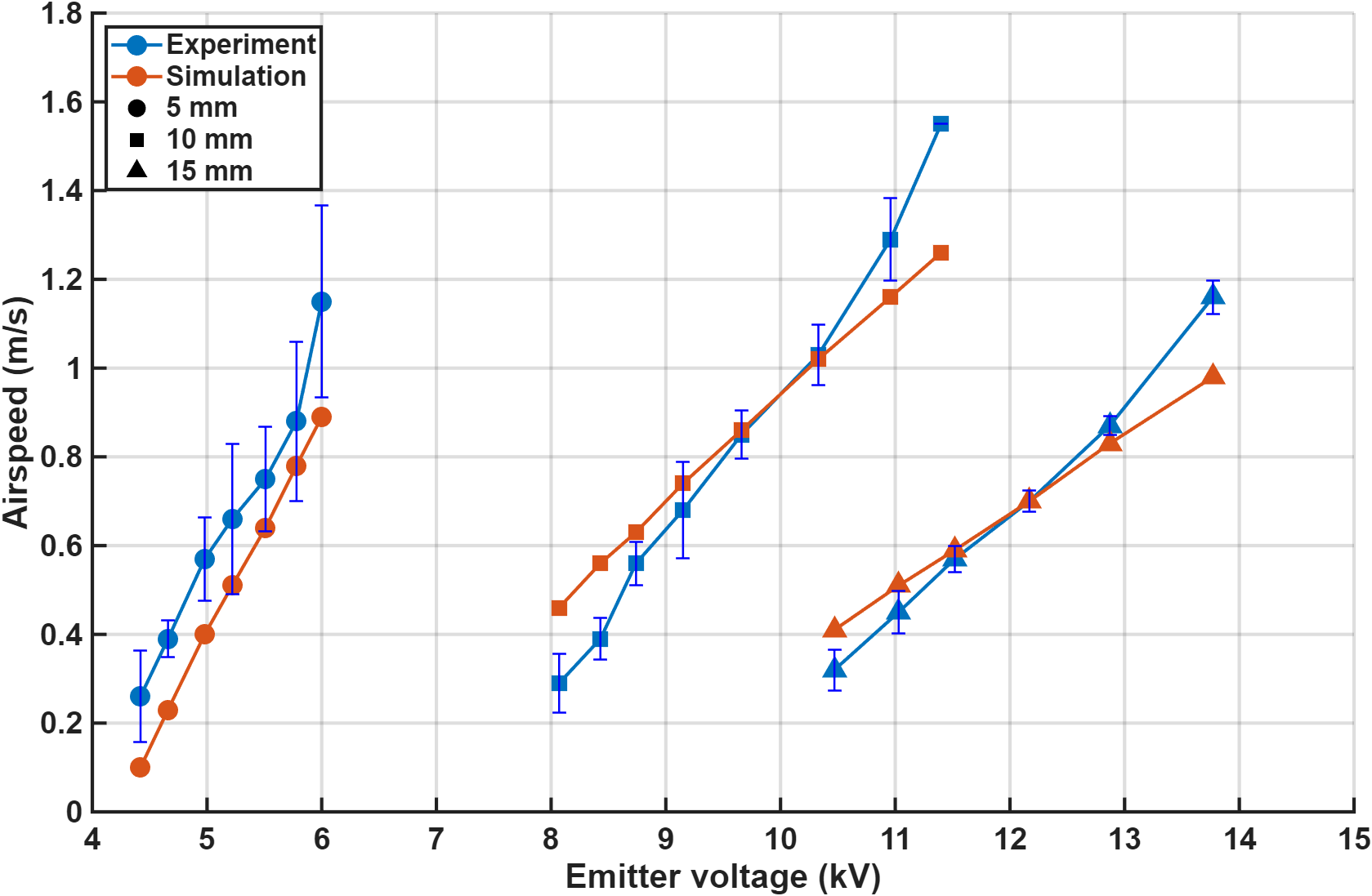}
        \caption{}
    \end{subfigure}
    \hfill
    \begin{subfigure}[b]{0.48\textwidth}
        \includegraphics[width=\textwidth]{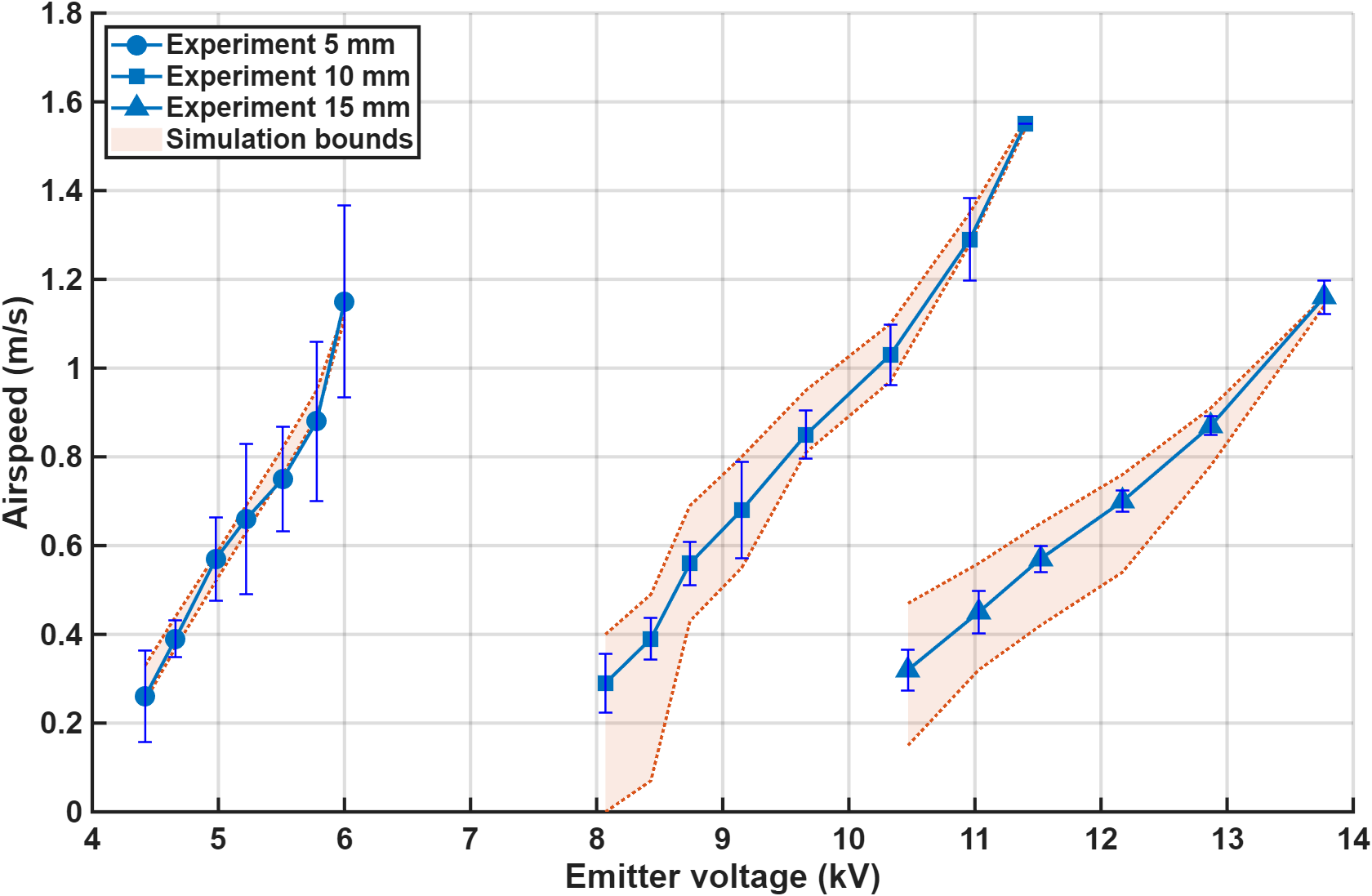}
        \caption{}
    \end{subfigure}
    \caption{Comparison of airspeeds between experiment and smooth surfaced emitter with 5, 10 and 15 mm gaps}
    \label{fig:f16}
\end{figure}

One key assumption in the simulations is that the emitter wire surface is perfectly smooth. SEM imaging of the Nichrome emitter wire (Fig. 7) reveals the presence of longitudinal micro-scale protrusions with characteristic heights of approximately 2 µm. Such surface features are known to locally enhance the electric field, thereby modifying corona inception conditions and increasing the kinetic energy of ions in the near-electrode region. These effects are not explicitly captured by the standard threshold inception formulation.

To assess the impact of surface morphology, representative longitudinal protrusions with a height of 2 µm were incorporated into the emitter geometry in the numerical model (Fig. 9). The protrusions locally increase the electric-field intensity at their tips, leading to earlier inception and enhanced charge injection at discrete locations along the emitter surface. At the same time, the concentration of ionization around these high-curvature regions reduces the effective ionization area compared to the idealized smooth-wire case.

To quantify the sensitivity of the system to surface morphology, a small variation in the number of protrusions was introduced. The shaded corridors shown in Fig. 8b represent simulations in which the number of protrusions along the emitter was varied by ±1 relative to the nominal configuration corresponding to the SEM-observed surface morphology. For each electrode gap, the upper and lower bounds therefore reflect the change in predicted airspeed resulting solely from this minimal geometric perturbation.

Physically, this corridor emerges because modifying the number of protrusions alters the spatial distribution of electric-field enhancement and, consequently, the total injected space charge. Even a single additional protrusion redistributes the ionization zones and modifies the integrated electrohydrodynamic body force acting on the flow. The resulting variation in body-force magnitude propagates through the coupled drift–diffusion–Poisson–Navier–Stokes system and appears as a bounded variation in outlet airspeed.

The sensitivity to protrusion count depends strongly on electrode spacing. For the 5 mm gap, the ionization region is already compact and intense, and the effective ionization area is dominated by the overall high electric-field gradient. In this regime, small geometric variations have only a minor influence on the total force generation. In contrast, at 10 mm and 15 mm gaps—particularly at lower voltages—the ionization region is more spatially extended and less intense. Under these conditions, small changes in effective ionization area produce proportionally larger variations in total space charge and therefore in airflow velocity.

At higher applied voltages, the overall field strength increases and the ionization region becomes sufficiently robust that the relative influence of individual protrusions diminishes. Consequently, the simulation corridor narrows with increasing voltage, indicating reduced sensitivity to micro-scale geometric perturbations.

\begin{figure}[H]
    \centering
    \begin{subfigure}[b]{0.48\textwidth}
        \includegraphics[width=\textwidth]{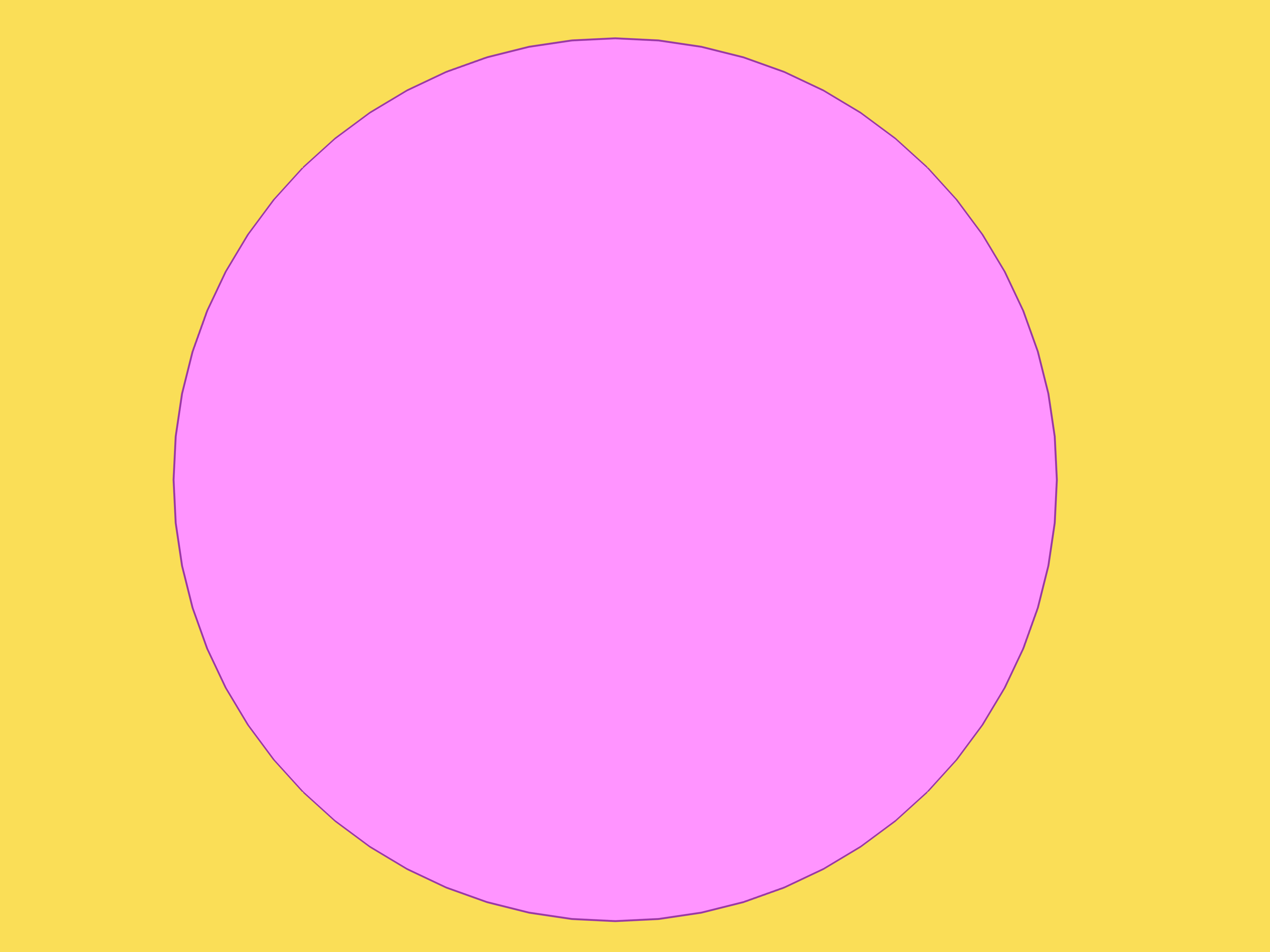}
        \label{fig:f11}
        \caption{}
    \end{subfigure}
    \hfill
    \begin{subfigure}[b]{0.48\textwidth}
        \includegraphics[width=\textwidth]{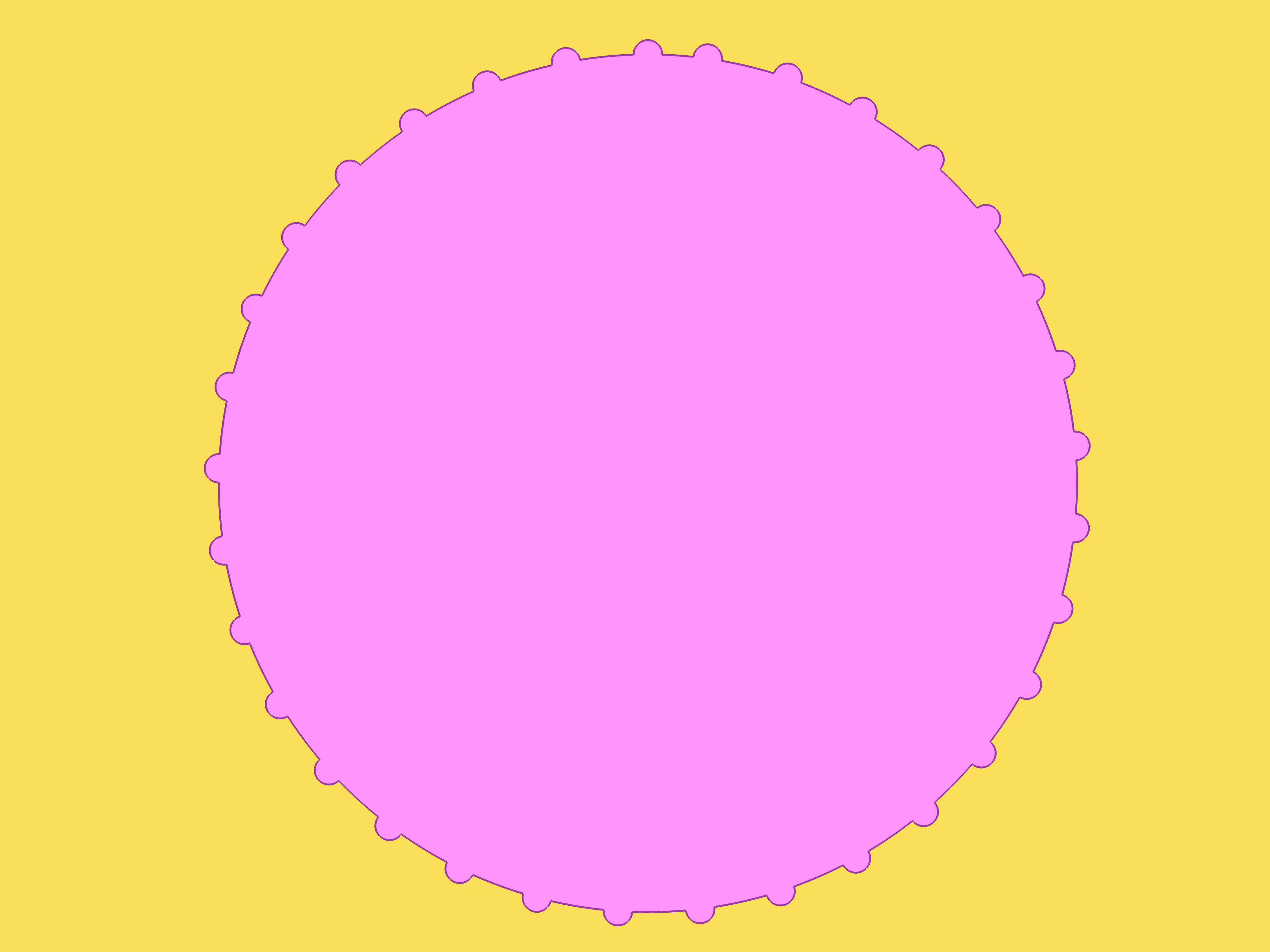}
        \label{fig:f12}
        \caption{}
    \end{subfigure}
    \caption{Cross-section views emitter wire with smooth surface (a) and longitudinal protrusions on the surface (b)}
    \label{fig:f10}
\end{figure}

Visualizations of the space charge density and electric-field distributions (Fig. 10) further illustrate the formation, structure, and spatial extent of the ionized regions in the vicinity of the emitter. The results show that ionization is highly localized near the emitter surface, with peak space charge densities occurring in regions of maximum electric-field intensity. For smaller electrode gaps and higher applied voltages, steeper electric-field gradients develop, leading to more compact and intense ionization zones adjacent to the emitter.

These localized high-gradient regions directly influence the magnitude and distribution of the electroaerodynamic body force, as the volumetric force density scales with the product of space charge density and electric field strength. Consequently, configurations exhibiting stronger near-emitter gradients generate higher local momentum transfer to the neutral flow, resulting in increased airflow velocities downstream. As the electrode gap increases, the ionized region becomes more spatially extended but less intense, which is reflected in the reduced force density and lower sensitivity of airflow generation to further voltage increases.

From an engineering standpoint, these results highlight that ionic wind performance is governed not only by the applied voltage level but also by the spatial organization of the ionization region. Small geometric or surface-induced variations that modify local electric-field enhancement can significantly alter the effective force distribution, even when global operating parameters remain unchanged. This observation reinforces the importance of accurately resolving near-electrode electric-field and charge-density gradients in predictive simulations intended for device design and optimization.

\begin{figure}[H]
    \centering
    \includegraphics[width=1\linewidth]{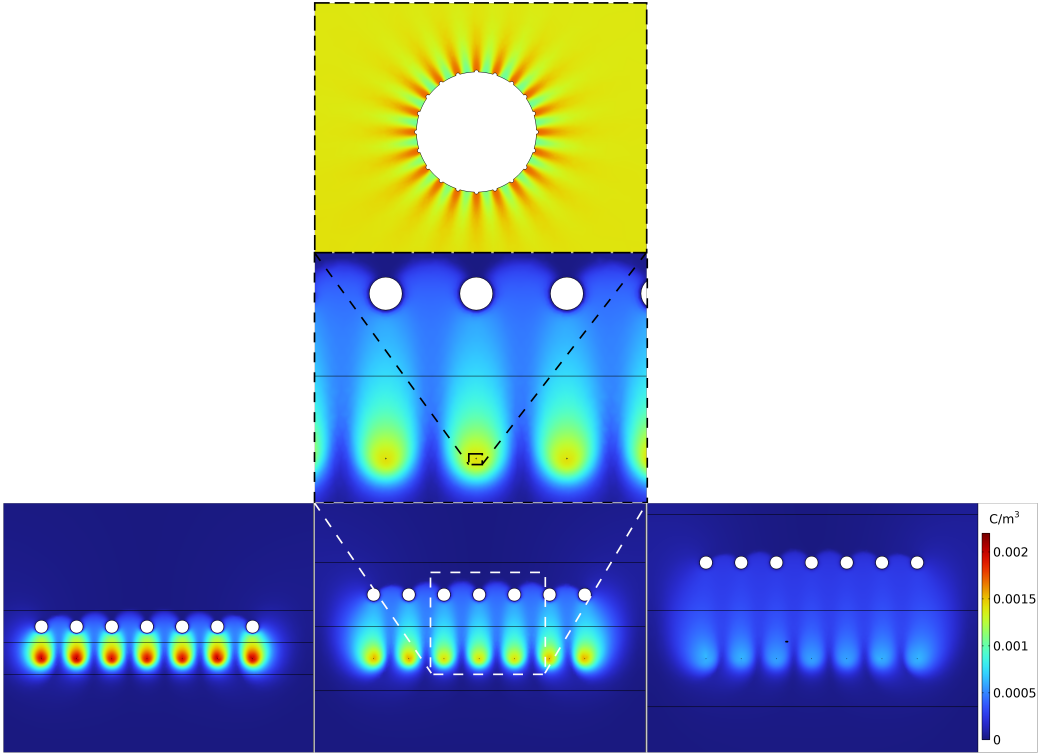}
    \caption{Charge density at different electrode gaps and distribution around the electrode}
    \label{fig:f11}
\end{figure}

\section{Conclusion}

This study evaluated the applicability of a threshold inception–based multiphysics modeling approach as an engineering-oriented predictive tool for ionic wind fans, using a wire-to-cylinder configuration as a controlled reference system. Direct comparison between simulations and experiments demonstrates that the model reliably captures the dominant electroaerodynamic mechanisms and reproduces key performance trends across varying electrode spacings and operating voltages. This confirms the suitability of the approach as a baseline framework for practical EAD device analysis.

From an engineering perspective, the validation process highlights an important limitation of commonly adopted modeling assumptions. Simulations based on idealized, perfectly smooth emitter surfaces systematically underestimate airflow generation at higher voltages and larger electrode gaps. Experimental characterization of the emitter wire shows that realistic micro-scale surface roughness is inherently present in manufactured electrodes and leads to localized electric-field enhancement, modified corona inception behavior, and increased momentum transfer to the neutral flow. Incorporating representative surface features into the numerical model significantly improves quantitative agreement with experimental measurements, particularly in regimes relevant for device scaling and higher-performance operation.

Rather than proposing a finalized correction model, this work demonstrates that electrode surface morphology constitutes a non-negligible design and modeling parameter that directly affects predictive accuracy. Neglecting such effects can limit the usefulness of numerical simulations for engineering design studies, especially when extrapolating beyond narrowly calibrated experimental conditions. The results therefore identify a clear pathway for improving the robustness and reliability of threshold inception–based models when transitioning from idealized laboratory configurations to practical devices.

Overall, the study establishes that threshold inception–based electroaerodynamic modeling, when carefully validated and augmented with physically motivated representations of electrode morphology, can serve as a reliable predictive tool for ionic wind fan performance. This capability is essential for advancing the technological readiness of EAD systems, enabling informed design trade-offs, geometry optimization, and scaling studies without excessive reliance on empirical tuning. The findings directly support the development of engineering-grade simulation methodologies for applications including airflow generation, active cooling, electrostatic air handling, and surface-integrated aerodynamic flow control.

By systematically linking corona discharge physics, electrode morphology, and macroscopic flow generation, this work contributes toward the development of reliable predictive tools for ionic wind devices. Such tools are essential for advancing the technological readiness of EAD systems and enabling informed design, optimization, and scaling studies for applications including airflow generation, active cooling, electrostatic air handling, and surface-integrated aerodynamic flow control.

Overall, the work highlights that micro-scale electrode morphology is a critical factor of EAD performance, and its inclusion in multiphysics simulations is essential for developing reliable predictive tools for practical applications such as electrostatic air filtration, active cooling, and aerodynamic drag reduction.

\section{Acknowledgements}
This work was supported by Estonian Research Council grants SEKMO 2021-2027.1.01.24-0730, PRG2675, TARISTU24-TK10, TEM-TA23 and COVSG31.

\section{Declaration of generative AI and AI-assisted technologies in the manuscript preparation process}

During the preparation of this work the author(s) used ChatGPT in order to improve readability. After using this tool, the author(s) reviewed and edited the content as needed and take(s) full responsibility for the content of the published article.

\bibliographystyle{ieeetr}
\bibliography{references}

\end{document}